\documentclass[10pt]{article}
\usepackage[margin=20mm]{geometry}

\usepackage{graphicx} 
\usepackage{amsmath}
\usepackage{natbib}
\usepackage{url}

\usepackage{amsthm}
\theoremstyle{plain}

\title{Bayesian Parameter Estimation of Normal Distribution from Sample Mean and Extreme Values}
\author{
Tomoki Matsumoto \thanks{Corresponding author: T.Matsumoto, Dr. of Engineering,
   Assistant Professor (Specially Appointed), School of Economics, University of Toyama,
    3190 Gofuku, Toyama City, Toyama, 930-8555, Japan,
    E-mail: t.matsumoto514@gmail.com, mtomoki@eco.u-toyama.ac.jp, 
    16-digit ORCID ID: 0000-0002-5680-7431} 
}

\begin{document}
\bibliographystyle{apalike}

\maketitle

\begin{abstract}
This paper proposes a Bayesian method for estimating the parameters of a normal distribution when only limited summary statistics (sample mean, minimum, maximum, and sample size) are available. 
To estimate the parameters of a normal distribution, we introduce a data augmentation approach using the Gibbs sampler, where intermediate values are treated as missing values and samples from a truncated normal distribution conditional on the observed sample mean, minimum, and maximum values. 
Through simulation studies, we demonstrate that our method achieves estimation accuracy comparable to theoretical expectations.

\vspace{3mm}
\end{abstract}


\section{Introduction}
In some survey reports, especially government surveys, raw data are often unavailable, and only summary statistics (sample mean, minimum, maximum, and sample size) are reported, whereas other statistics, such as sample variance, are not included. 
For example, in Japan, \cite{kato2012waste}, \cite{mhlw2016welfare}, and \cite{cabinet2023npo} followed this reporting format.
Although this limited information restricts the direct application of many statistical techniques, it still provides useful information for estimating the parameters of a normal distribution.
Previous work has addressed parameter estimation of a normal distribution from limited information: 
Direct estimation formulas for means and standard deviations from summary statistics (\cite{Hozo2005-ov}) and estimation methods using only the maximum values (\cite{Capaldi2019-xr}).

This paper proposes a Bayesian estimation based on data augmentation through the Gibbs sampler that leverages available summary statistics to estimate the parameters of a normal distribution. 
The proposed method treats intermediate values, except for the minimum and maximum values, as missing values and samples from a truncated normal distribution conditional on the observed sample mean, minimum, and maximum to achieve data augmentation. 
This augmentation enables us to model the data as if a complete dataset were available.
Our approach provides a computationally feasible solution using a Gibbs sampler while maintaining estimation accuracy comparable to theoretical expectations.

\section{Method}

Consider samples $x = \{ x_1, x_2, \ldots, x_n \}$ where each $x_i \sim \text{N}(\mu, \sigma^2)$ independently, and define the order statistics $\{ x_{(1)}, x_{(2)}, \ldots, x_{(n)} \}$ where $x_{(1)}$ is the minimum value and $x_{(n)}$ is the maximum value in $x$. 
The complete samples are unavailable, but only the sample mean $\bar{x}$, minimum $x_{(1)}$, maximum $x_{(n)}$, and sample size used to calculate these summary statistics are available. 
Our objective is to estimate the parameters of the normal distribution, $\mu$ and $\sigma^2$, from these limited summary statistics.

\subsection{Gibbs Sampler with Data Augmentation}
We treat the intermediate values $x_{(2)}, \ldots, x_{(n-1)}$ as missing values, and assume that they are independent samples from a truncated normal distribution, given the observed values $x_{(1)}$ and $x_{(n)}$, with bounds defined by $x_{(1)} \le x_{(j)} \le x_{(n)}$ for $j=2, \ldots, n-1$.
Although the order statistics are generally not independent, conditioning on $x_{(1)}$ and $x_{(n)}$ allows us to treat the intermediate values as independent within these truncated bounds. 
This is feasible by considering any bijective ordering function $\text{Order}(x)$ that maps the original sample values $x_{j^{'}}$ to the order statistics $x_{(j)}$. Given the range $x_{(1)} \le x_{(j)} \le x_{(n)}$, we have $x_{(1)} \le \text{Order}^{-1}(x_{(j)}) = x_{j^{'}} \le x_{(n)}$.
In our setting, because we are primarily interested in the values of intermediate samples rather than their exact ordering, we can identify an order statistic $x_{(j)} = \text{Order}(x_{j^{'}})$ using the original sample $x_{j^{'}}$ and treat the intermediate values as independent samples for parameter estimation of a normal distribution. 
After augmenting the intermediate values using samples from a truncated normal distribution, the sample set that includes the augmented values $x_{(j)}^{\ast}$ ($j=2, \ldots, n-1$), $x_{(1)}$ and $x_{(n)}$ can be regarded as independent samples from $\text{N}(\mu, \sigma^{2})$ before ordering.

A truncated normal distribution is a modified version of the normal distribution, where the values are restricted to lie within a specified range $[a, b]$ (\cite{Robert1995-ub}). 
The probability density function (PDF) of a truncated normal distribution $\text{TN}(x \mid \mu, \sigma^2, a, b)$ is defined as
\begin{align*}
\text{TN}(x \mid \mu, \sigma^2, a, b) = \frac{\phi(x \mid \mu, \sigma^{2})}{\Phi(b \mid \mu, \sigma^{2}) - \Phi(a \mid \mu, \sigma^{2})}, \quad a \leq x \leq b,    
\end{align*}
\noindent where $\phi(x | \mu, \sigma^2)$ and $\Phi(x | \mu, \sigma^2)$ are the PDF and cumulative distribution function (CDF) of the normal distribution with mean $\mu$ and variance $\sigma^2$, respectively.
The truncated normal distribution ensures that all the generated values remain within the known minimum and maximum bounds.
However, the mean of a truncated normal distribution is basically not equal to the mean of the original normal distribution.

To modify the bias above, we define the adjusted mean \(\bar{x}_{\text{adj}}\), which separates the influence of the minimum and maximum on the mean as follows:
\begin{align*}
\bar{x}_{\text{adj}} = \frac{1}{n - 2}\left(n\bar{x} - x_{(1)} - x_{(n)}\right).    
\end{align*}
\noindent This adjusted mean accounts for only the intermediate values.
To ensure that the mean of the augmented dataset equals the observed sample mean $\bar{x}$, we first center the boundaries by subtracting $\bar{x}_{\text{adj}}$: $a^{\ast} = x_{(1)} - \bar{x}_{\text{adj}}$ and $b^{\ast} = x_{(n)} - \bar{x}_{\text{adj}}$.
We then adjust the mean parameter of the truncated normal distribution with the range $[a^{\ast}, b^{\ast}]$ so that the expected value of the sampled data equals zero.
This parameter can be derived from the properties of the truncated normal distribution: 
If $Z \sim \text{TN}(\mu, \sigma^2, a^{\ast}, b^{\ast})$, then
\begin{align*}
E[Z] = \mu + \sigma\frac{\phi(a^{\ast} \mid \mu, \sigma^{2}) - \phi(b^{\ast} \mid \mu, \sigma^{2})}{\Phi(b^{\ast} \mid \mu, \sigma^{2}) - \Phi(a^{\ast} \mid \mu, \sigma^{2})}.    
\end{align*}
Setting $E[Z] = 0$ and solving for $\mu$ yields the mean parameter required for sampling with zero mean samples. 
We define it as $\mu_{\text{trunc}}$, the second term of the right-hand side times $-1$:
\begin{align*}
    \mu_{\text{trunc}} = -\sigma\frac{\phi(a^{\ast} \mid \mu, \sigma^{2}) - \phi(b^{\ast} \mid \mu, \sigma^{2})}{\Phi(b^{\ast} \mid \mu, \sigma^{2}) - \Phi(a^{\ast} \mid \mu, \sigma^{2})}.
\end{align*}
Finally, after sampling $z_{(j)}^{\ast}$, $j=1,\ldots,n-1$ from $\text{TN}(\mu_{\text{trunc}}, \sigma^{2}, a^{\ast}, b^{\ast})$, we add back $\bar{x}_{\text{adj}}$ to the sampled values, that is, $x_{(j)}^{\ast} = z_{(j)}^{\ast} + \bar{x}_{\text{adj}}$.
This ensures that the mean of $x^{\ast} = \{x_{(1)}^{\ast}, x_{(2)}^{\ast}, \ldots, x_{(n-1)}^{\ast}, x_{(n)}^{\ast}\}$, where $x_{(1)}^{\ast} = x_{(1)}$ and $x_{(n)}^{\ast} = x_{(n)}$, is $E[\bar{x}^{\ast}] = E[\bar{x}] = \mu$ while maintaining the constraints $x_{(1)} \leq x_{(j)}^{\ast} \leq x_{(n)}$.

To construct an efficient Gibbs sampler for our problem, we use the following conditional conjugate prior distributions to the normal distribution with the mean parameter $\mu$ and variance parameter $\sigma^{2}$: 
For mean $\mu$, the prior distribution is a normal distribution with mean $\mu_{0}$ and variance $\tau_{0}^{2}$;
For variance $\sigma^{2}$, the prior distribution is the following inverse gamma distribution:
\begin{align*}
\text{InvGamma}\left(\sigma^{2} \mid \alpha_0, \beta_0\right) = \frac{\beta_{0}^{\alpha_{0}}}{\Gamma(\alpha_{0})} (\sigma^2)^{-(\alpha_{0}+1)} \exp\left(-\frac{\beta_{0}}{\sigma^2}\right).
\end{align*}
Their conjugacy enables straightforward sampling from the posterior distributions through the Gibbs sampler (\cite{BDA3}).
We propose the following Gibbs sampler with data augmentation.
 
\begin{itemize}
    \item \textbf{Step 1}: Initialize $\mu^{(0)}$ and $\sigma^{2(0)}$.\footnote{In the following, we omit the superscripts for simplicity.}
    \item \textbf{Step 2}: Sample the missing values from a truncated normal distribution:
    \[
    z_{(j)}^{\ast} \sim \text{TN}(\mu_{\text{trunc}}, \sigma^2, a^{\ast}, b^{\ast}), ~ j=2,\ldots,n-1,
    \]
    and after transforming $z_{(j)}^{\ast}$ into $x_{(j)}^{\ast}$, set $x^{\ast} = \{x_{(1)}^{\ast} = x_{(1)}, x_{(2)}^{\ast}, \ldots, x_{(n-1)}^{\ast}, x_{(n)}^{\ast} = x_{(n)}\}$.
    \item \textbf{Step 3}: Sample \(\mu\) using the conditional posterior:
    \[
    \mu \mid \sigma^2, \bar{x} \sim \text{N}\left(\frac{\mu_0 / \tau_0^2 + n\bar{x} / \sigma^2}{1 / \tau_0^2 + n / \sigma^2}, \frac{1}{1 / \tau_0^2 + n / \sigma^2}\right)
    \]
    \item \textbf{Step 4}: Sample \(\sigma^2\) using the inverse gamma posterior:
    \[
    \sigma^2 \mid \mu, x^{\ast} \sim \text{InvGamma}\left(\alpha_0 + \frac{n}{2}, \beta_0 + \frac{1}{2}\sum_{i=1}^n (x_{(i)}^{\ast} - \mu)^2\right)
    \]
    \item \textbf{Step 5}: Repeat steps 2-4 until convergence.
\end{itemize}

\section{Simulation Study and Results}
We conducted simulations to assess the performance of our Gibbs sampler with data augmentation. 
Datasets were generated from $\text{N}(0, 5^{2})$ for different sample sizes, $N = \{10, 50, 100, 500, 1000\}$.
To examine robustness, we tested some prior configurations, each defined by $\mu_0$, $\tau^2$, $\alpha_0$, and $\beta_0$.
The detailed settings are shown in the figures below.
Each simulation involved $10,000$ iterations, where the first $5,000$ iterations were discarded as the burn-in period.

\subsubsection*{Single Simulation}
\begin{figure}[ht]
    \centering
    \includegraphics[width=1\linewidth]{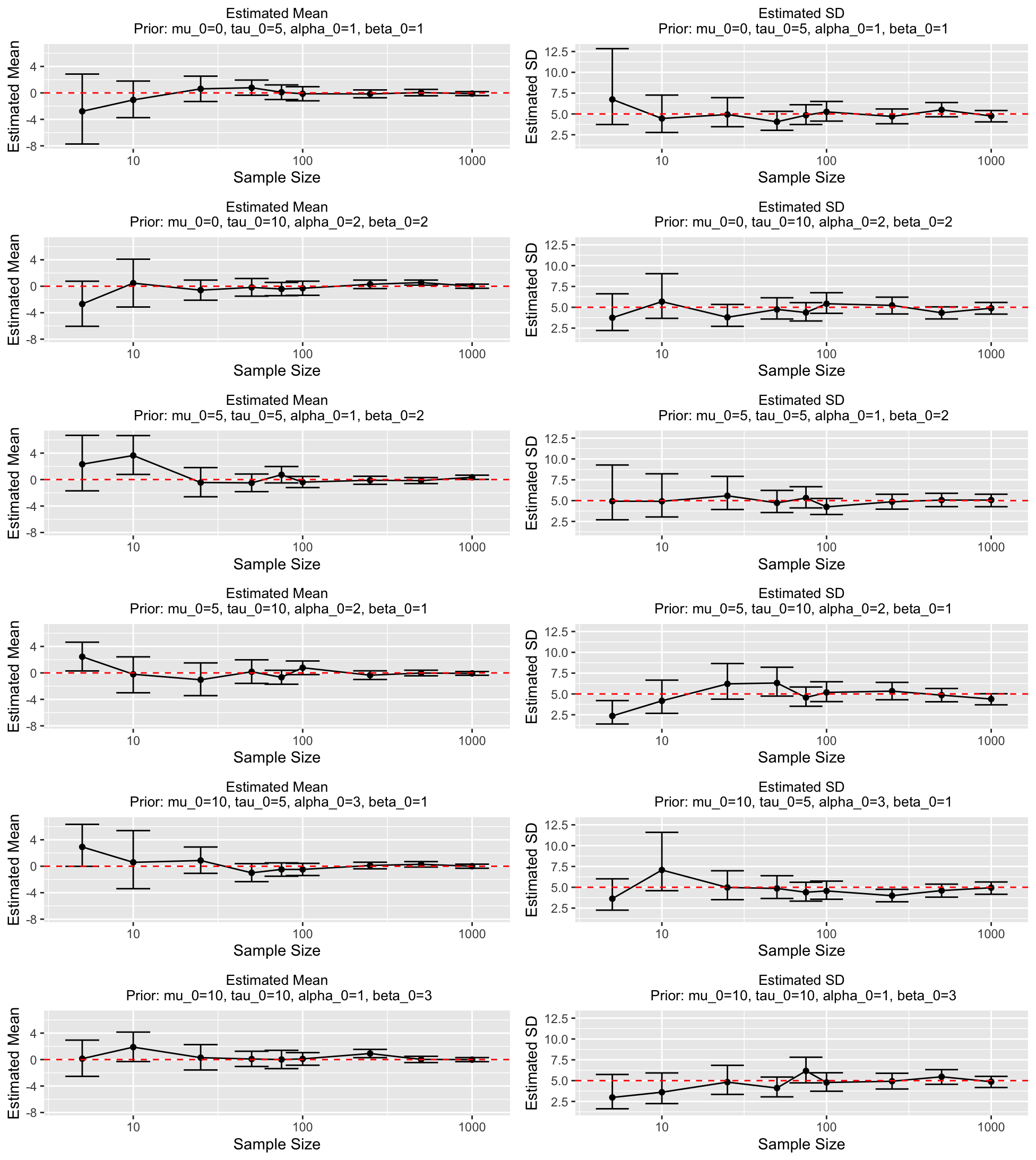}
    \caption{Posterior estimates across different sample sizes. 
    Black dots represent posterior means with their corresponding 95\% credible intervals shown as vertical bars. 
    The red dotted line indicates the true values ($\mu = 0$ and $\sigma = 5$).
    The horizontal axis is shown on a logarithmic scale.}
    \label{fig:sigle_result}
\end{figure}

We performed a single simulation for each sample size.
Figure \ref{fig:sigle_result} shows that both posterior estimates for the mean parameter $\mu$ and the variance parameter $\sigma^2$ improved accuracy with larger sample sizes.
Additionally, reasonable estimates were obtained even with small sample sizes.
The results suggest that both parameters can be reliably estimated even with moderate sample sizes.

\subsubsection*{Comparison with Theoretical Distributions of the Summary Statistics}

The theoretical distributions of the minimum and maximum for a normal distribution with mean $\mu$ and variance $\sigma^{2}$ are as follows:
\begin{align*}
f_{X_{(1)}}(x \mid \mu, \sigma^{2}) &= n \cdot \left(1 - \Phi\left(x \mid \mu, \sigma^{2}\right)\right)^{n-1} \cdot \phi\left(x \mid \mu, \sigma^{2}\right),\\
f_{X_{(n)}}(x \mid \mu, \sigma^{2}) &= n \cdot \Phi\left(x \mid \mu, \sigma^{2}\right)^{n-1} \cdot \phi\left(x \mid \mu, \sigma^{2}\right).
\end{align*}
To derive the likelihood function $L$, we consider the adjusted mean $\bar{X}_{\text{adj}}$ to separate the influence of the minimum and maximum on the mean.
The expected value of the adjusted mean can be easily calculated as $\mu$.\footnote{Because the expected deviation from the mean for the minimum and maximum values, \(E[(X_{(1)} - \mu) + (X_{(n)} - \mu)]\), is zero due to the symmetry of the normal distribution, $E[X_{(1)} + X_{(n)}] = 2\mu$.
Thus, the expected value of the adjusted mean is $E[\bar{X}_{\text{adj}}] = \left(n\mu - 2\mu\right) / (n - 2) = \mu$.}
On the other hand, the correct calculation of the variance of the adjusted mean is hard because of the remaining correlations among $\bar{X}_{\text{adj}}$, $X_{(1)}$, and $X_{(n)}$ (\cite{OrderStat}).
Thus, using a truncated normal given $X_{(1)}$ and $X_{(n)}$ on $\bar{X}_{\text{adj}}$ will be more practical:
\begin{align}
L(\mu, \sigma^2 \mid \bar{x}, x_{(1)}, x_{(n)}) = \text{TN}\left(\bar{x}_{\text{adj}} \mid \mu, \frac{\sigma^2}{n-2}, x_{(1)}, x_{(n)} \right) \cdot f_{X_{(1)}}(x_{(1)} \mid \mu, \sigma^2) \cdot f_{X_{(n)}}(x_{(n)} \mid \mu, \sigma^2).
\label{eq:theoretical_likelihood}
\end{align}
However, this likelihood function cannot be expressed in closed form.

\begin{figure}[ht]
    \centering
    \includegraphics[width=1\linewidth]{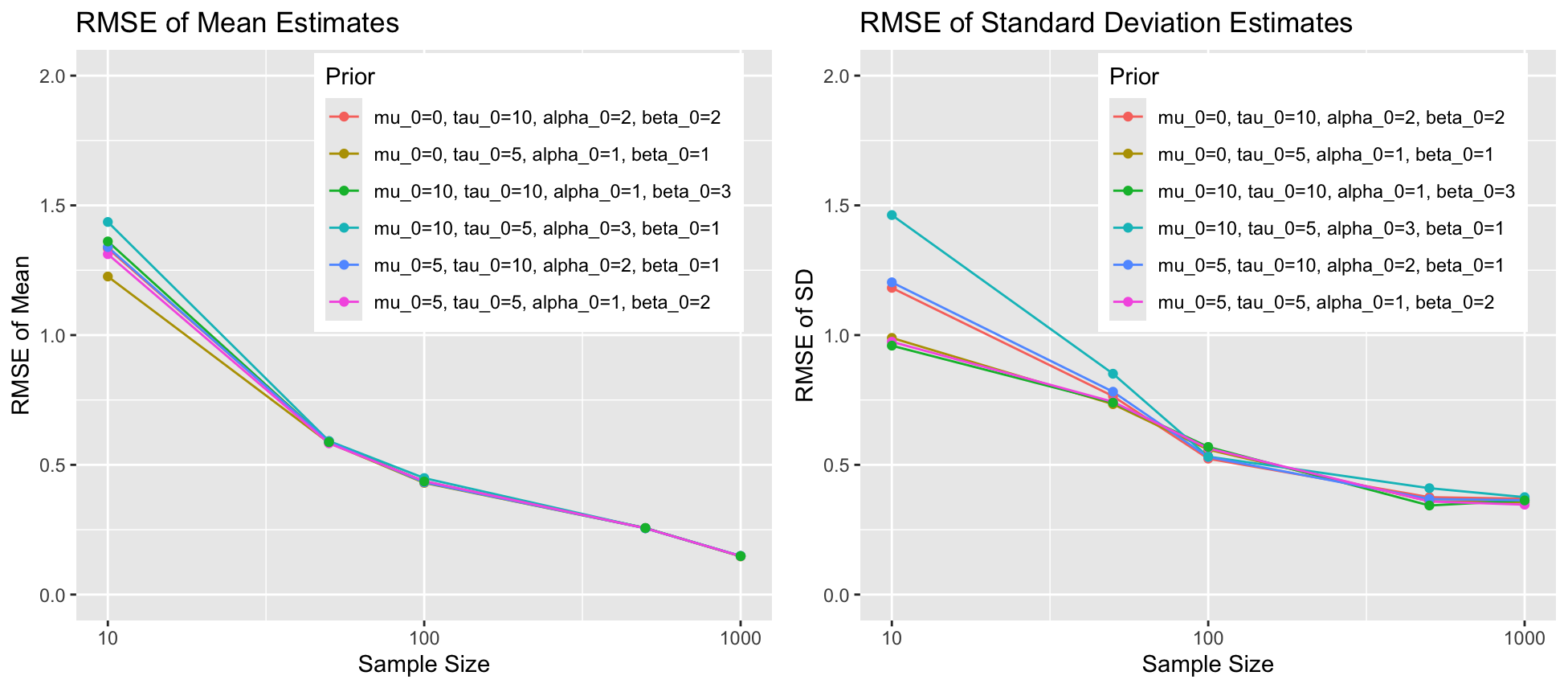}
    \includegraphics[width=1\linewidth]{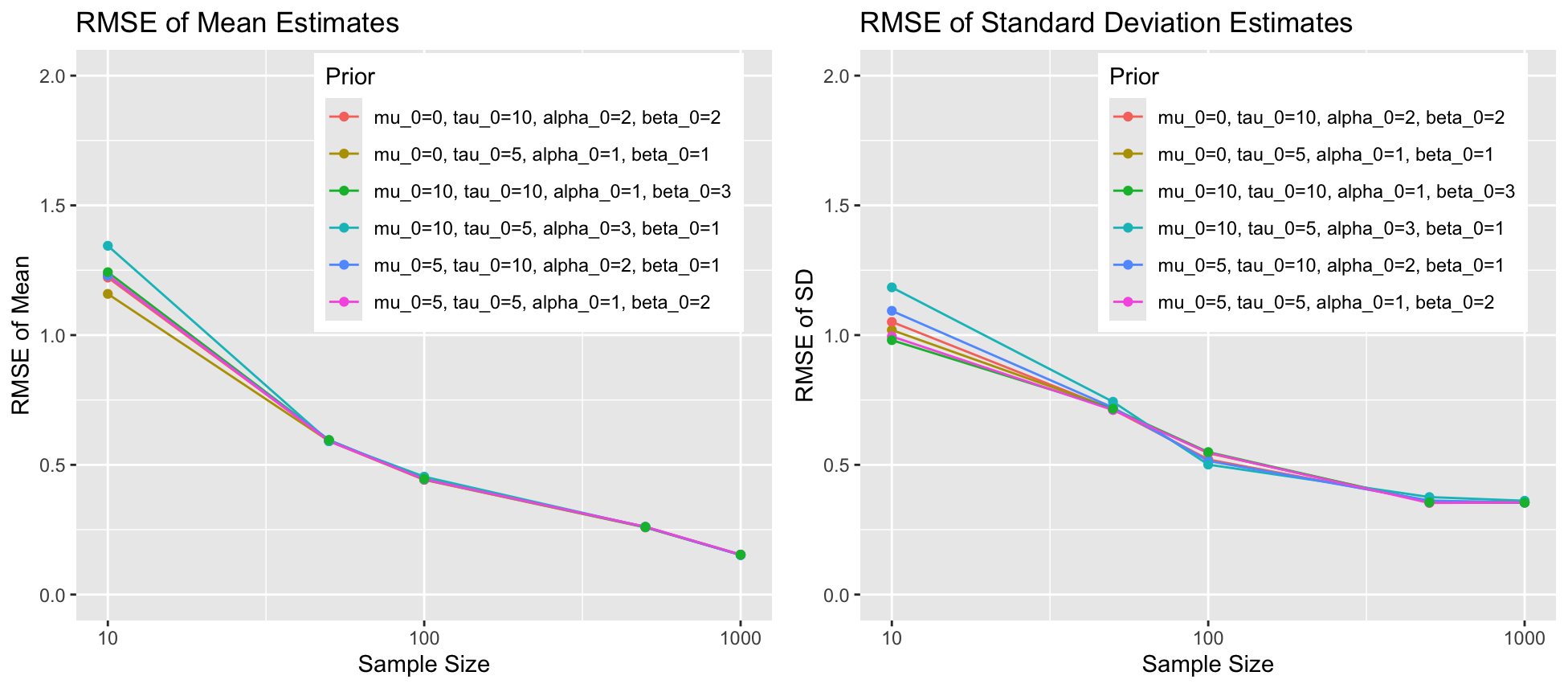}
    \caption{RMSE comparison between the proposed method and theoretical distributions. 
    Results from 20 simulations for each sample size.
    Top: Results from the proposed Gibbs sampling method.
    Bottom: Results from the theoretical distribution using RStan.
    The horizontal axis is shown on a logarithmic scale.}
    \label{fig:comparison_results}
\end{figure}

To validate the effectiveness of the proposed method, we conducted a comparative analysis with the theoretical considerations using RStan (\cite{RStan}) to generate posterior samples based on the likelihood (\ref{eq:theoretical_likelihood}). 
Figure \ref{fig:comparison_results} shows the Root Mean Squared Error (RMSE) calculated from 20 simulations for each sample size. 
The comparison demonstrated that our method's estimates closely aligned with theoretical expectations, and larger sample sizes improved the estimation accuracy for both approaches.
Additionally, reasonable estimates were obtained, even with small sample sizes, particularly when the parameters of the prior distributions were appropriately specified.
The proposed method offers a practical approach that achieves accuracy comparable to theoretical expectations.
While the theoretical approach requires more complex sampling methods, such as Metropolis Hastings or Hamiltonian Monte Carlo, due to the non-closed form expression of the model, our Gibbs sampling approach enables straightforward implementation through direct sampling from conditional posterior distributions by using conjugate priors and data augmentation.

\section{Conclusion}
This paper proposed a Bayesian approach using Gibbs sampling to estimate the parameters of a normal distribution from limited summary statistics.
Our method provides a practical solution for parameter estimation when only the summary statistics are available.
Simulation results demonstrated that the proposed method provides stable and reliable parameter estimates.
Our data augmentation approach with Gibbs sampling offers a practical and computationally efficient solution while maintaining accuracy comparable to theoretical expectations.


\bibliography{ref}
\end{document}